# Scaling of Berry-curvature monopole dominated large linear positive magnetoresistance

Shen Zhang,[1,2,*] Yibo Wang,[1,2,*] Qingqi Zeng,[1] Jianlei Shen,[1] Xinqi Zheng,[3] Jinying Yang,[1,2] Zhaosheng Wang,[4] Chuanying Xi,[4] Binbin Wang,[1] Min Zhou,[5] Rongjin Huang,[5] Hongxiang Wei,[1] Yuan Yao,[1] Shouguo Wang,[3] Stuart S. P. Parkin,[6] Claudia Felser,[7†] Enke Liu,[1,8†] Baogen Shen[1,9†]

[1]Beijing National Laboratory for Condensed Matter Physics, Institute of Physics, Chinese Academy of Sciences, Beijing 100190, China

[2]School of Physical Sciences, University of Chinese Academy of Sciences, Beijing 100049, China

[3]School of Materials Science and Engineering, Beijing Advanced Innovation Center for Materials Genome Engineering, University of Science and Technology Beijing, Beijing 100083, China

[4]Anhui Province Key Laboratory of Condensed Matter Physics at Extreme Conditions, High Magnetic Field Laboratory, Chinese Academy of Sciences, Hefei 230031, China

[5]Key Laboratory of Cryogenics, Technical Institute of Physics and Chemistry, Chinese Academy of Sciences, Beijing 100190, China.

[6]Max Planck Institute of Microstructure Physics, D-06120 Halle, Germany

[7]Max Planck Institute for Chemical Physics of Solids, 01187 Dresden, Germany

[8]Songshan Lake Materials Laboratory, Dongguan 523808, China

[9]Ningbo Institute of Materials Technology & Engineering, Chinese Academy of Sciences, Ningbo, Zhejiang, 315201, China

**The linear positive magnetoresistance (LPMR) is a widely observed phenomenon in topological materials, which is promising for potential applications on topological spintronics. However, its mechanism remains ambiguous yet and the effect is thus uncontrollable. Here, we report a quantitative scaling model that correlates the LPMR with the Berry curvature, based on a ferromagnetic Weyl semimetal $CoS_2$ that bears the largest LPMR of over 500% at 2 Kelvin and 9 Tesla, among known magnetic topological semimetals. In this system, masses of Weyl nodes existing near the Fermi level, revealed by theoretical calculations, serve as Berry-curvature monopoles and low-effective-mass carriers. Based on the Weyl picture, we propose a relation $\mathrm{MR} = \frac{e}{\hbar} B \Omega_F$, with $B$ being the applied magnetic field and $\Omega_F$ the average Berry curvature near the Fermi surface, and further introduce temperature factor to both MR/$B$ slope (MR per unit field) and anomalous Hall conductivity, which establishes the connection between the model and experimental measurements. A clear picture of the linearly slowing down of carriers, i.e., the LPMR effect, is demonstrated under the cooperation of the $k$-space Berry curvature and real-space magnetic field. Our study not only provides an experimental evidence of Berry curvature induced LPMR for the first time, but also promotes the common understanding and functional designing of the large Berry-curvature MR in topological Dirac/Weyl systems for magnetic sensing or information storage.**

In the past decade, the development of topological materials has intensively brought about exotic physical phenomena and much novel understanding of physical matters[1-4]. Among these materials, the study of topological Weyl materials has always been related to the Berry curvature[5-7]. Berry curvature originates from the electronic transition between adjacent energy levels in the parameter space[8]. In particular, the $k$-space Berry curvature is related to band crossing in electronic energy band structures, such as the Weyl node, which is exactly the point of focus in the research of topological materials. However, the Berry curvature would be zero if the material possesses protection from time-reversal and space-inversion symmetries[9]. Weyl nodes can be treated as $k$-space magnetic monopoles with strong Berry curvature ($k$-space pseudo-magnetic field) around Weyl nodes. Thus, Weyl systems, or Dirac systems under magnetic fields, are perfect platforms for studying non-zero Berry curvatures.

Studies on the effect of Berry curvature on transport properties in topological materials have generally focused on chiral anomaly[10, 11] and transverse transport phenomena such as the intrinsic anomalous Hall effect[11-13] and the intrinsic anomalous Nernst effect[14-16]. Berry curvature is believed to generate the intrinsic component in these anomalous transverse transport effects. Based on theoretical models, the anomalous Hall conductivity is the integration of the entire Berry curvature under the Fermi surface, and the anomalous transverse thermoelectric conductivity is the integration of the Berry curvature around the Fermi surface[17].

Meanwhile, linear positive magnetoresistance (LPMR) has also been widely reported in topological materials, including Dirac semimetals[18-21], Weyl semimetals[22-24], and magnetic Weyl semimetals[25-27]. The large magnetoresistance could be potentially used in magnetic field sensors, disk reading heads and magnetic memory[28-31]. However, explanations for LPMR are ambiguous till now. Abrikosov proposed a mechanism called quantum magnetoresistance, which is determined by extrinsic impurity scattering and requires all electrons to be filled in one Landau band[32]. Parish and LittleWood presented that a macroscopically disordered material can exhibit the LPMR based on a random network model[33]. Feng et al proposed that the LPMR may arise from the splitting process from Dirac cones to Weyl nodes in magnetic fields[34]. Song et al proposed a semiclassical theory that guiding center diffusion could cause the LPMR[35]. Zhang et al reported that the LPMR can be obtained by a two-Weyl-node model with inter-node scattering[36, 37]. Imran and Hershfield mentioned that a slight linear change in MR is caused by the modification of phase space with magnetic fields[38]. These mechanisms made contributions to understand the LPMR behavior. However, it is hard to experimentally control the parameters in these models. There thus still lacks experimental confirmation for these mechanisms of large LPMRs in topological materials[39-42].

Recently, the pyrite $CoS_2$ was reported as a magnetic Weyl semimetal candidate[43, 44]. In this study, we found that $CoS_2$ bears many Weyl nodes and a large LPMR. We proposed an intrinsic model showing a physical picture in which the LPMR is determined by the Weyl-node Berry curvature around the Fermi surface. And the temperature-dependent anomalous Hall conductivity and LPMR obtained in experiments can be well fitted by using this intrinsic model. This is the first time that the experimental evidence of the relation between the LPMR and Berry curvature has been reported. This model can further work well in reported topological semimetals with LPMRs, and can help to discover new materials with a large non-saturating linear magnetoresistance.

$CoS_2$ single crystals were prepared using a flux method (see Section 1 in Supplementary Information). The crystal has a cubic structure (Fig. 1a), with a space group of Pa-3 (No. 205). We measured the powder X-ray diffraction spectrums from 5 to 300 K and refined them by the Fullprof software (Fig. 1b). The lattice parameter decreases as the temperature decreases, which is well consistent with the moving trend of peaks. But below 25 K, we found a negative thermal expansion of the unit cell (see Section 2 in Supplementary Information). The Co atom centers at an octahedron consisting of six nearest neighbor S atoms, and each S atom is shared by three octahedrons. A single Co atom layer in the *xy* plane perfectly shows the space relation of these octahedrons (Fig. 1c). There are two different types of Co-S octahedrons that lie in two different directions. These two types interlace with each other, which can be confirmed by high-resolution electron microscopy (Fig. 1d).

Our magnetization measurements indicated that $CoS_2$ is an excellent soft ferromagnet, with a Curie temperature of 124 K (Fig. 1e). In a low field, the sharp magnetic transition around $T_C$, the stable magnetization below $T_C$, and the slight difference between ZFC and FC curves, indicate that a strong and pure ferromagnetic interaction was established in the high-quality crystals. Meanwhile, a negligible difference between the magnetization curves in different directions (***B*** // [100], [110], [111], see Section 3 in Supplementary Information) shows a weak magnetocrystalline anisotropy in $CoS_2$, although its easy magnetization axis was determined as [111] in our study and a previous report[45].

The longitudinal resistivity of $CoS_2$ decreases with decreasing temperature in almost the entire temperature range from 2 to 300 K, except for a hump around the Curie temperature (Fig. 1g). This hump was addressed using electron correlation theory in the year 1980[46]. Below the Curie temperature, a rapid decrease in longitudinal resistivity is observed, leaving a quite low residual resistivity of 0.56 μΩ cm at 2 K. This value is much lower than that of many magnetic metallic systems, including magnetic Weyl semimetal $Co_3Sn_2S_2$ (50 μΩ cm)[11]. Meanwhile, a high residual resistivity ratio (RRR, $\rho_{300\ K}/\rho_{2\ K}$) of 247 was observed (Fig. 1e). Both results show a strong indication of high conductivity and high quality in $CoS_2$ crystals.

Our theoretical calculations showed linear band crossing structures in the electronic energy bands (Fig. 2a). The calculated band structures on the R-X and M-Γ paths, and the small electron pocket around R point, are in agreement with the reported angle-resolved photoemission spectroscopy (ARPES) results[43]. The differences in the energy bands at 5 and 30 K coincided with the transition of lattice parameters revealed by XRD (see Section 2 in Supplementary Information). We further found many Weyl fermions in this material. Within the range of 10 meV around the Fermi level, there are eight pairs of Weyl nodes with opposite chirality near M points in the Brillouin zone (Fig. 2b), among which four pairs of Weyl nodes are just located at the Fermi level (see Section 4 in Supplementary Information). We further observed more linear-dispersion band crossings with strong Berry curvature in wider energy (±30 meV) ranges, and their influence on the transport properties is remarkable because they are close to the Fermi energy, similar to the case of $Co_3Sn_2S_2$ with the Weyl nodes 60 meV above the Fermi level[11]. Benefitting from the masses of Weyl nodes and linear-dispersion band structures, the theoretically calculated anomalous Hall conductivity (AHC) achieves approximately 900 $\Omega^{-1}$ cm$^{-1}$, and an AHC peak appears at the energy point, that is, 10 meV lower than the Fermi energy (Fig. 2c). Four apparent Weyl nodes at the $k_y$-$k_z$ plane ($k_x = 0$) can perfectly represent the Weyl-node-dominated Fermi surface in $CoS_2$ (Fig. 2d). Strong Berry curvatures lie in a Weyl node and remain stable around the Fermi energy (Fig. 2e). The Berry curvature at Weyl nodes achieves a spectacular $10^4$ Bohr$^2$ (Fig. 2f), which significantly We measured the field-dependent magnetoresistance and Hall resistances at different temperatures as the magnetic field along the [001] direction and current along the [100] direction (inset of Fig. 3a). Below 30 K, the Hall conductivity is dominated by a normal Hall effect (Fig. 3a), with a typical Hall response of a two-carrier system. With increasing temperature, the characteristic of the anomalous Hall effect appears at 15 K, which is obvious in the Hall resistivity curves (Fig. 3b). The reason why the anomalous Hall resistivity seems to vanish below 15 K is that the signal-to-noise ratio in measurement is not high enough. This phenomenon was also reported in other materials[25, 26]. Notably, the maximal value of Hall conductivity, 50,000 $\Omega^{-1}$ cm$^{-1}$, is achieved at 2 K and 4 T. This super high Hall conductivity is not related to the anomalous Hall effect but originates from the high longitudinal conductivity, which reaches $1.8 \times 10^6$ $\Omega^{-1}$ cm$^{-1}$ calculated from the low residual resistivity of 0.56 μΩ cm (see Fig. 1e).

Focusing on the field-dependent Hall resistivity with increasing temperature, we observed a transition from a two-carrier to a single-carrier character (Fig. 3b). By applying semi-classical two-carrier model[47] on the normal Hall effect part of Hall resistivity curves, we extracted the carrier concentration ($n_h$~$2.3 \times 10^{20}$ cm$^{-3}$, $n_e$~$1.2 \times 10^{22}$ cm$^{-3}$) and carrier mobility ($\mu_h$~6,100 cm$^2$ V$^{-1}$ s$^{-1}$, $\mu_e$~820 cm$^2$ V$^{-1}$ s$^{-1}$) at 2 K. When the temperature increases to approximately 25 K, the hole mobility decreases by over one order of magnitude; thus, the contribution of holes to electrical transport becomes faint, and the normal Hall resistivity becomes to confirm to the single-carrier model (Fig. 3c). Additionally, from both the carrier concentration and mobility, we can conclude that the high conductivity in $CoS_2$ comes from two aspects: large numbers of electrons and high-mobility holes. The high-mobility holes are considered to be related to the Weyl fermions. The existence of high-mobility holes is consistent with the results of our calculations that Weyl nodes are near the Fermi energy.

The anomalous Hall resistivity (AHR) can be extracted from linear extrapolation of the high-field Hall resistivity curve. The AHC is then calculated by

$$\sigma_{xy}^A = \frac{\rho_{yx}^A}{\left(\rho_{yx}^A\right)^2 + \rho_{xx}^2} \qquad (1)$$

Here, $\rho_{yx}^A$ is the AHR at zero field, and $\rho_{xx}$ is the longitudinal resistivity at zero field.

The AHC is dominated by the intrinsic mechanism (see Section 5 in Supplementary Information). It changes slightly below 30 K, and decreases rapidly with increasing temperature above 30 K (Fig. 3d), which will be explained later by the Berry curvature model. The difference between the measured (700 $\Omega^{-1}$ cm$^{-1}$) and calculated (900 $\Omega^{-1}$ cm$^{-1}$) values is caused by a slight cobalt deficiency confirmed by chemical and XRD analysis in real materials (see Sections 1 and 2 in Supplementary Information). The cobalt deficiency causes a lower Fermi energy of about −20 meV, where the AHC is calculated to be 700 $\Omega^{-1}$ cm$^{-1}$ (Fig. 2c). Nevertheless, the Fermi energy in real materials is still close to the calculated Weyl nodes, and the small Fermi pocket at the R point observed in previous ARPES measurement[43] is also retained.

Large linear positive magnetoresistance (LPMR) is observed in CoS$_2$ below 25 K (Fig. 3e). At 2 K and 7 T, MR reaches more than 400%, while the MR curves above 20 K are significantly small to observe, as their absolute values are less than 10% in the present magnetic field. When the magnetic field increases to 32 T, the positive MR does not show any signature of saturation (see Section 6 in Supplementary Information). In contrast to the simulated MR using the two-carrier model along with the carrier concentration and mobility extracted from the Hall resistivity curves, the measured MR data is far deviated from the two-carrier model (Fig. 3f). This means that a notable mechanism dominates the LPMR in the current system.

The explanations for LPMR are multifarious and there has not been a consensus regarding it until now. The quantum magnetoresistance[32] requires the system approaches the quantum limit and the carrier concentration needs to be very small (< $10^{18}$ cm$^{-3}$), which is hard to be satisfied by a real semimetal (> $10^{18}$ cm$^{-3}$). For CoS$_2$, the carrier concentration is ~$10^{22}$ cm$^{-3}$ (Fig. 3c), implying that this system cannot be properly described by the quantum magnetoresistance model. The random network model[33] is based on the strongly inhomogeneous distribution of atoms, which is not suitable for high-quality crystals like the current CoS$_2$ (Fig. 1e). The guiding center theory[35, 40] requires the Hall angle to be independent of the magnetic field and be about 1, but the Hall angle of CoS$_2$ is magnetic field dependent and much less than 1 (see Section 6 in Supplementary Information). So, it is hard to apply existing models, especially extrinsic mechanisms, to explain the LPMR in CoS$_2$.

Thus, we proposed a Berry-curvature mechanism. Based on the Berry-curvature-modified motion equations of Bloch electrons[9], the electron velocity, longitudinal conductivity and the anomalous Hall conductivity are given by

$$\dot{\boldsymbol{r}} = \frac{1}{\hbar}\nabla_k \varepsilon + \frac{e}{\hbar}\boldsymbol{E}\times\boldsymbol{\Omega} + \frac{e}{\hbar}(\dot{\boldsymbol{r}}\times\boldsymbol{B})\times\boldsymbol{\Omega} \quad (2)$$

$$\sigma_{xx} = -e^2\tau\int\frac{d\boldsymbol{k}}{(2\pi)^3}\frac{\partial f_0}{\partial \varepsilon}\frac{1}{D(\boldsymbol{k})}v_x^2 \quad (3)$$

$$\sigma_{yx}^{A} = \frac{e^2}{\hbar}\int\frac{d\boldsymbol{k}}{(2\pi)^3}\Omega_z f_0 \quad (4)$$

Here, $D(\boldsymbol{k})=1+\frac{e}{\hbar}\boldsymbol{B}\cdot\boldsymbol{\Omega}$, and we assumed that the electric field is perpendicular to the magnetic field. Then we made an approximation for the Berry curvature in Eq. (3), and the MR was deduced as

$$\text{MR} = \frac{e}{\hbar}B\Omega_F \quad (5)$$

The details of the deduction process can be found in the Section 7 in Supplementary Information. The Berry curvature in Eq. (5) represents the average value of the Berry curvature near the Fermi surface. The LPMR will be higher if the Berry curvature near the Fermi surface is larger.

A schematic illustration of the carrier motion affected by Berry curvature and magnetic field can visually exhibit our model (Fig. 4a). Once an external magnetic field is applied to the system, the carriers will experience a Lorentz force. For a carrier with a strong Berry curvature in the momentum space, this force subsequently produces an opposite velocity (the third term in Eq. (2)) that is antiparallel to the initial one, which in turn slows down the motion of carriers, leading to an increase in the MR. For topologically nontrivial electronic bands such as Weyl nodes, we can expect exceptionally high-velocity fermions and strong Berry curvature compared to those of normal bands. Therefore, the increasing resistivity induced by the deceleration of these carriers can be prominent and large. Based on the theoretical calculations and Hall measurements (Figs. 2 and 3), a mass of Weyl nodes exists around the Fermi surface in $CoS_2$, which generates many high-mobility carriers and large Berry curvatures near the Fermi surface. Thus, $CoS_2$ has a prominent longitudinal conductivity and positive MR among magnetic topological semimetals (Fig. 4d).

To confirm our model, we considered a simple three-dimensional model with only two Weyl nodes near the Fermi surface (see details in the Section 7 in Supplementary Information). The Hamiltonian, energy bands and the Berry curvature of the Weyl node at (0, 0, $c$) are

$$H = \lambda[k_x\sigma_x + k_y\sigma_y - (k_z - c)\sigma_z] + \Delta \quad (6)$$

$$\varepsilon_\pm = \Delta \pm \lambda\sqrt{k_x^2 + k_y^2 + (k_z - c)^2} \quad (7)$$

$$\boldsymbol{\Omega}_\pm(\boldsymbol{k}) = \pm \frac{1}{2\sqrt{k_x^2 + k_y^2 + (k_z - c)^2}^3}[k_x, k_y, k_z - c], \quad (8)$$

where $\Delta$ represents the shift from the Fermi energy, and $\lambda$ represents the dispersion slope. Then we got the temperature-dependent AHC and LPMR induced by the Berry curvature, which are given by

$$\sigma_H^A(T) = \frac{e^2}{\hbar}\int d\boldsymbol{k}\Omega_z f_0 = \frac{e^2}{4\pi^2\hbar}(2c - \frac{|\Delta|}{2\lambda} - \frac{k_BT}{\lambda}e^{-\frac{|\Delta|}{2k_BT}}) \quad (9)$$

$$\frac{MR}{B}(T) = \frac{e}{\hbar}\Omega_F(T) = \frac{e}{\hbar}\frac{\int d\boldsymbol{k}\Omega_z\partial_\varepsilon f_0}{\int d\boldsymbol{k}\partial_\varepsilon f_0} = \frac{e\lambda^2}{4\hbar\Delta^2}\frac{\tanh\left(\frac{|\Delta|}{2k_BT}\right)}{1 + \frac{\pi^2}{3}\left(\frac{k_BT}{\Delta}\right)^2} \quad (10)$$

These two equations are important in this study as they relate the theoretical model and experimental data. From Eq. (9) we can know that, the AHC is positively correlated with $2c$, the distance between two Weyl nodes. This is consistent with the known theoretical understanding about the intrinsic anomalous Hall effect[48, 49], in which the AHC is determined by the separation of two Weyl nodes in magnetic Weyl semimetals. Besides, it can be seen from Eq. (9), the AHC is further negatively correlated with $\Delta$, the shift of Weyl nodes from the Fermi energy. We then collected the AHC data of $Co_3Sn_2S_2$[11], $Co_2MnGa$[13] and $CoS_2$, as they are magnetic Weyl systems and exhibit large AHC. The temperature-dependent AHC can be well fitted by Eq. (9) (Fig. 4b). The extracted shift value from the Fermi energy of $CoS_2$ is about 10 meV, which is highly consistent with our calculations in this work. In addition, the AHC of $CoS_2$ decreases above 30 K, and the AHC of $Co_3Sn_2S_2$ decreases above 100 K. This difference originates from the difference of the shift from the Fermi energy, as the extracted $|\Delta|$ of $Co_3Sn_2S_2$ is about 80 meV, approaching the reported value of 60 meV[11]. The AHC would start decreasing at a higher temperature if the shift from the Fermi energy is larger.

The experimental AHC data can be well fitted by using our model, which confirms the validity of this model and motivates us to do the fitting of the LPMR data obtained in our and other experiments. We extracted the slopes of the MR curves below 100 K for $CoS_2$, $NbP$[24], $ZrSiS$[21], $FeP$[20], $PrAlSi$[27], and $MnBi$[25], as they exhibit LPMR. We applied Eq. (10) to these data and produced well fitted results, as shown in Fig. 4c, which demonstrates that our intrinsic model also works well for longitudinal transport LPMR. Therefore, we provided the evidence of Berry curvature induced LPMR by fitting the temperature-dependent AHC and LPMR in topological materials by using our intrinsic model. The LPMR observed in Dirac, Weyl, and magnetic Weyl systems is believed to be connected to the Berry curvature of Weyl nodes.

In this work, we demonstrated $CoS_2$ is a magnetic Weyl semimetal with masses of Weyl fermions (Figs. 2 and 3). How to understand the connection between the Berry curvature and LPMR in other topological materials? In topological systems, the linear crossing band structure, where the Weyl node lies, can produce a strong Berry curvature around it. In Dirac systems, a pair of Weyl nodes with opposite chirality lie at the same coordinate in $k$-space, forming a Dirac node, as required by time-reversal and space-inversion symmetries. Therefore, the Berry curvature is zero everywhere in the $k$-space. In Weyl systems, with broken space-inversion symmetry, Dirac nodes are split into Weyl nodes, resulting in non-zero Berry curvature around the Weyl nodes. However, as required by the time-reversal symmetry, at the opposite coordinates lie two Weyl nodes with the same chirality, and the generated Berry curvatures have opposite signs. Therefore, the integral of the Berry curvature in the $k$-space is zero. In magnetic Weyl systems with broken time-reversal symmetry, both the Berry curvature and the integral of the Berry curvature are non-zero. Broken time-reversal symmetry is necessary to observe the influence of the Berry curvature. This is not a problem with the MR measurements, because the application of a magnetic field can also break the time-reversal symmetry. Therefore, the LPMR can be expected in topological semimetal systems based on intrinsic model.

However, LPMR is not observed in all topological materials, even if these materials generally contain a large Berry curvature. This is because the influence of the Berry curvature on fermions can be covered by other mechanisms, especially when trivial band structures are located at the Fermi level; for example, the MR in magnetic Weyl metal $Co_2MnGa$ is small and negative, and is mainly dominated by high-concentration trivial electrons and magnetic scattering[50, 51]. Meanwhile, the LPMR is not contained in the whole magnetic field region. Nearly quadratic MR at low magnetic fields and a slight deviation from the linear relation at high magnetic fields were observed in $CoS_2$ (see Section 6 in Supplementary Information) and other systems[19, 23, 25-27]. These phenomena may result from the influence of the magnetic field on the Berry curvature or band splitting.

Our model is a semiclassical model with a concise Berry curvature assumption, ignoring the possible trivial band structures and the influence of magnetic field on relaxation time and electron energy. And we also considered the difference between single- and multiple-Weyl nodes in modelling (see Section 7 in Supplementary Information). This intrinsic model requires further research in the future, especially in the case of complex material systems. Even so, the present system $CoS_2$ provides an ideal platform to uncover the Berry curvature mechanism on the LPMR behavior that widely exists in magnetic or nonmagnetic topological semimetals.

In summary, large linear positive MRs are extensively contained in emerging topological materials in applied magnetic fields. The large linear positive magnetoresistance effect was studied based on the magnetic Weyl semimetal $CoS_2$. Benefitting from the large numbers of Weyl nodes and strong Berry curvature around the Fermi surface, $CoS_2$ offers the first example exhibiting the highest conductivity and the largest positive MR among known magnetic topological materials. To well understand the LPMR behavior in $CoS_2$, we proposed an intrinsic model in which the slope of the linear MR is determined by the Berry curvature near the Fermi surface and reveals the interplay effect of the real-space magnetic field and $k$-space Berry curvature on the LPMR behavior. And we provided experimental evidence of the connection between LPMR and Berry curvature by fitting the temperature-dependent AHC and LPMR. This intrinsic mechanism is further expected to operate in Dirac semimetals, Weyl semimetals, and other materials with strong Berry curvatures. Our study motivates unveiling the LPMR of topological systems and promotes the design of large LPMR materials for spintronics applications.

**Materials and Methods**

**Single-crystal growth.** The single crystals of $CoS_2$ were grown by a flux method in a $Al_2O_3$ crucible sealed in a quartz tube. The quartz tube was heated to 800 °C in 12 hours and kept there for 24 hours, then slowly cooled to 600 °C in 7 days. The chemical compositions of crystals were measured by Inductively Coupled Plasma-Atomic Emission Spectrometry (ICP-AES).

**Structural characterization.** The lattice parameter and S atom position are refined from powder X-ray diffraction spectrums by the Fullprof software. The powder X-Ray Diffraction spectrums were measured from 5 to 300 K by using a Rigaku SmartLab X-ray diffractometer. The measurement temperature step below 30 K is 2 K, in order to reveal the details of lattice parameter and S atom position changing with temperature. High resolution HAADF images were acquired in JEOL ARM200F transmission electron microscope equipped with cold field-emission gun and double Cs correctors. The convergence angle of the probe beam was about 22 mrad and the acceptance angle of the detector was between 70 and 200 mrad.

**Magnetization measurements.** Magnetization measurements were carried out on $CoS_2$ single crystals with the magnetic field applied along [100], [110], [111] respectively using the Magnetic Property Measurement System (MPMS). The results show a negligible magnetocrystalline anisotropy.

**Electrical transport measurements.** The measurements on longitudinal and Hall resistivity were performed on the Physical Property Measurement System (PPMS) using the resistivity option. The measured samples were cut out of a thin cuboid. During electrical transport measurements, the magnetic field was perpendicular to the sample plane and the current was along long side. The high-field transport measurements were carried out in water-cooled magnet with the steady fields up to 35 T at the High Magnetic Field Laboratory of Chinese Academy of Sciences by using standard a.c. lock-in techniques.

**Two-carrier model analysis.** The Hall resistivity below 20 K shows nonlinear field dependence, indicating the existence of two types of carriers (electrons and holes). The applied two-carrier model equations[52] are

$$\rho_{xy}(B) = \frac{B}{e} \frac{\left(n_e\mu_e^2 - n_h\mu_h^2\right) + \left(n_e - n_h\right)\mu_e^2\mu_h^2 B^2}{\left(n_e\mu_e + n_h\mu_h\right)^2 + \left(n_e - n_h\right)^2\mu_e^2\mu_h^2 B^2} \qquad (11)$$

$$\rho_{xx}(B=0) = \frac{1}{e\left(n_e\mu_e + n_h\mu_h\right)} \qquad (12)$$

Here, $B$ is the applied magnetic field, $\rho_{xy}$ is the Hall resistivity, $\rho_{xx}$ is the longitudinal resistivity, $n_e$ is the electron concentration, $\mu_e$ is the electron mobility, $n_h$ is the hole concentration, $\mu_h$ is the hole mobility. Since the longitudinal resistivity shows large linear positive magnetoresistance instead of two-carrier character, we only applied zero-field two-carrier expression for longitudinal resistivity.

**Density functional theory (DFT) calculations.** DFT implemented in Vienna Ab initio Simulation Package (VASP)[53-55] was used in our calculation. For local spin-density approximation (LSDA) calculations, the exchange correlation term is described according to the projected augmented wave (PAW) pseudopotentials. For self-consistent calculations, a Monkhorst-Pack method of 9×9×9 was used. Then, we constructed Wannier maximally localized function for d orbitals on cobalt site and p orbitals on sulfur site to project onto the ab initio bands using Wannier90 program[56]. The projected bands were perfectly fitted to DFT calculated bands. Then we calculated Berry curvature on 2D-plain using Wannier90 program. Using the data calculated from Wannier90, we calculated energy-dependent anomalous Hall conductivity using Wannierberri code[57, 58]. Wanniertools[59] was used to calculate the Weyl nodes. We also used Vaspkit[60] for post-processing of band structures.

**Acknowledgments**

We thank Dr. Shunqing Shen and Dr. Hongming Weng for fruitful discussions. This work was supported by Fundamental Science Centre of the National Natural Science Foundation of China (No. 52088101), National Key R&D Program of China (No. 2019YFA0704900), Beijing Natural Science Foundation (No. Z190009), National Natural Science Foundation of China (Nos. 11974394, 12174426, 11874359), the Strategic Priority Research Program (B) of the Chinese Academy of Sciences (CAS) (XDB33000000), the Key Research Program of CAS (No. ZDRW-CN-2021-3), the Synergetic Extreme Condition User Facility (SECUF), and the Scientific Instrument Developing Project of CAS (No. ZDKYYQ20210003). C. Felser thanks CAS President's International Fellowship Initiative for Distinguished Scientists. A portion of this work was performed on the Steady High Magnetic Field Facilities, High Magnetic Field Laboratory, CAS, and supported by the High Magnetic Field Laboratory of Anhui Province.

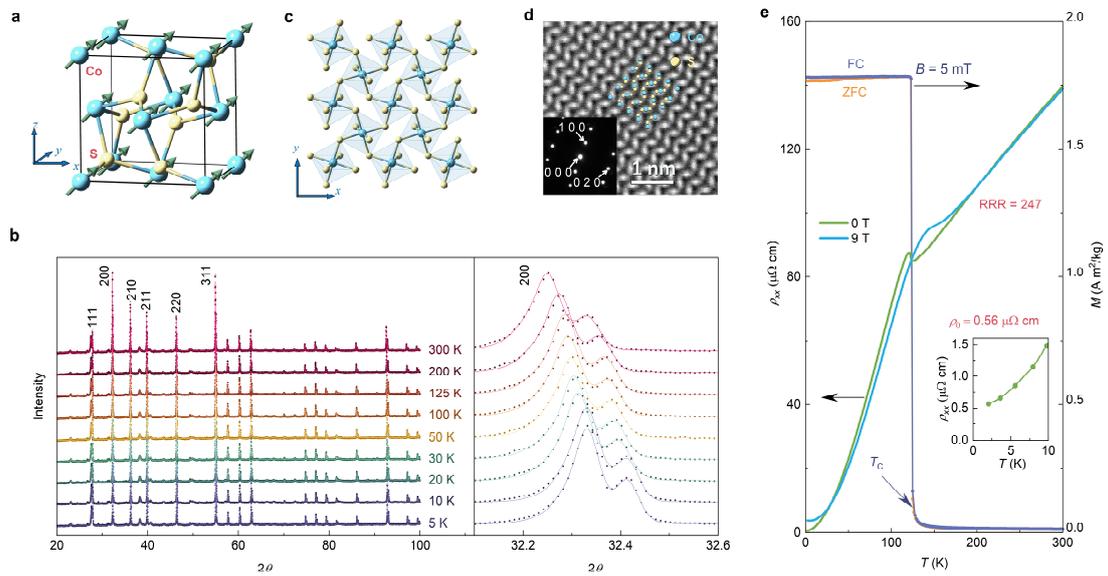

**Fig. 1 | Crystal structure and basic characterizations. a**, Unit cell. Co atoms form a face-centered cube. The easy magnetization axis is along [111]. **b**, Temperature-dependent powder X-ray diffraction spectrums and refinement curves. **c**, The spatial arrangement of a layer of Co-S octahedrons. **d**, Atomic resolution high-angle-annular-dark-field (HAADF) image of $CoS_2$ along the [001] axis, overlaid with the atom configuration and the electron diffraction patterns. **e**, Temperature dependencies of longitudinal resistivity in 0 and 9 T, and zero-field-cooling/field cooling (ZFC/FC) magnetization in 5 mT. The Curie temperature of $CoS_2$ is detected to be 124 K. In zero field, a residual resistivity of 0.56 µΩ cm and a residual resistivity ratio (RRR, $\rho_{300\,K}/\rho_{2\,K}$) of 247 were observed.

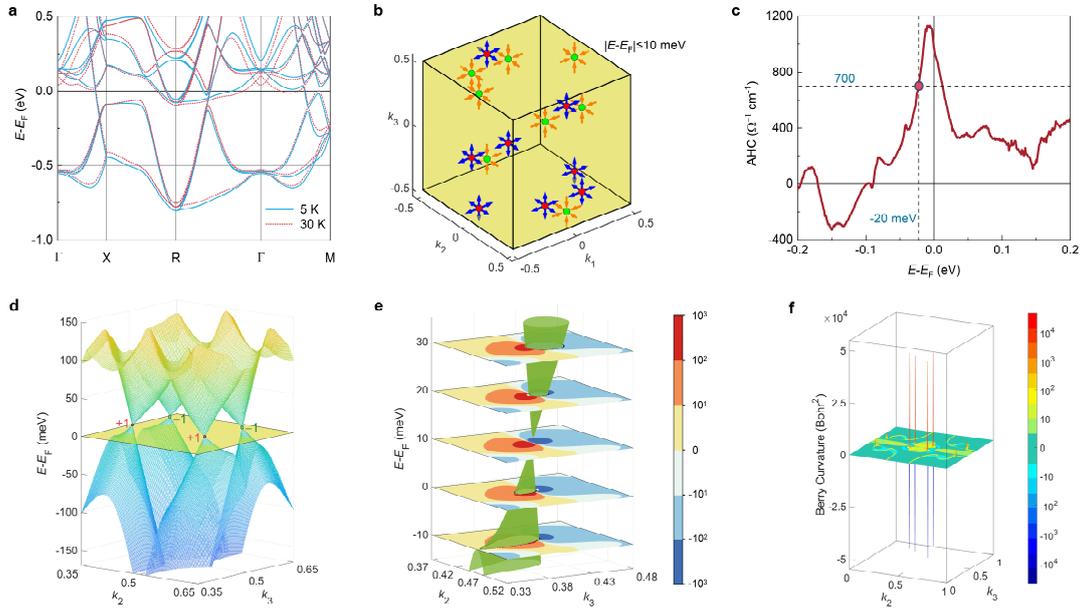

**Fig. 2 | Theoretical calculations of anomalous Hall conductivity and Berry curvature. a**, Energy bands calculated using the lattice parameters at 5 and 30 K. **b**, Eight pairs of Weyl nodes in Brillouin zone within the range of 10 meV around Fermi energy. **c**, Energy dependence of the anomalous Hall conductivity. The real material has a lower Fermi energy of about −20 meV according to the experimental AHC as 700 $\Omega^{-1}$ cm$^{-1}$. **d**, Two pairs of Weyl cones near M point in the Brillouin zone, 10 meV above the Fermi level. **e**, A Weyl node and the Berry curvatures around it. The color bar shows the value of Berry curvature in the unit of square Bohr. **f**, Berry curvature in $k_y$-$k_z$ plane ($k_x = 0$).

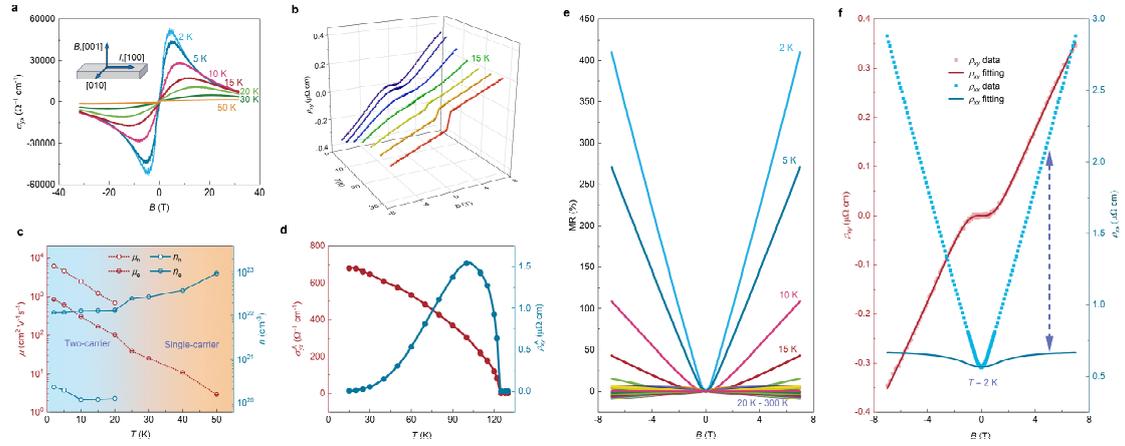

**Fig. 3 | Electrical transport signals**. **a**, High-field Hall conductivity. Normal-Hall-dominated Hall conductivity reaches 50,000 $\Omega^{-1}$ cm$^{-1}$ at 2 K and 4 T. The inset shows the directions of current and magnetic fields in the measurements. **b**, Field-dependent Hall resistivity below 30 K. When the temperature increases from 2 to 30 K, the normal Hall resistivity changes from two-carrier character to single-carrier character, and the anomalous Hall effect around low fields begins to be observable at 15 K and becomes clearer as the temperature increases. **c**, Carrier concentration and carrier mobility extracted from normal Hall resistivity from 2 to 50 K. In CoS$_2$, the concentration of electrons reaches $10^{22}$ cm$^{-3}$ and the mobility of holes reaches as high as 6,100 cm$^2$ V$^{-1}$ s$^{-1}$. **d**, Temperature dependence of anomalous Hall conductivity and anomalous Hall resistivity. The anomalous Hall conductivity achieves 700 $\Omega^{-1}$ cm$^{-1}$ at low temperatures. **e**, MR between ±7 T. Large LPMR is observed in CoS$_2$ below 20 K. **f**, Hall and longitudinal resistivity at 2 K and their fitting curves by the two-carrier model. The Hall resistivity can be well fitted by the two-carrier model, but the MR is dominated by a strong LPMR mechanism.

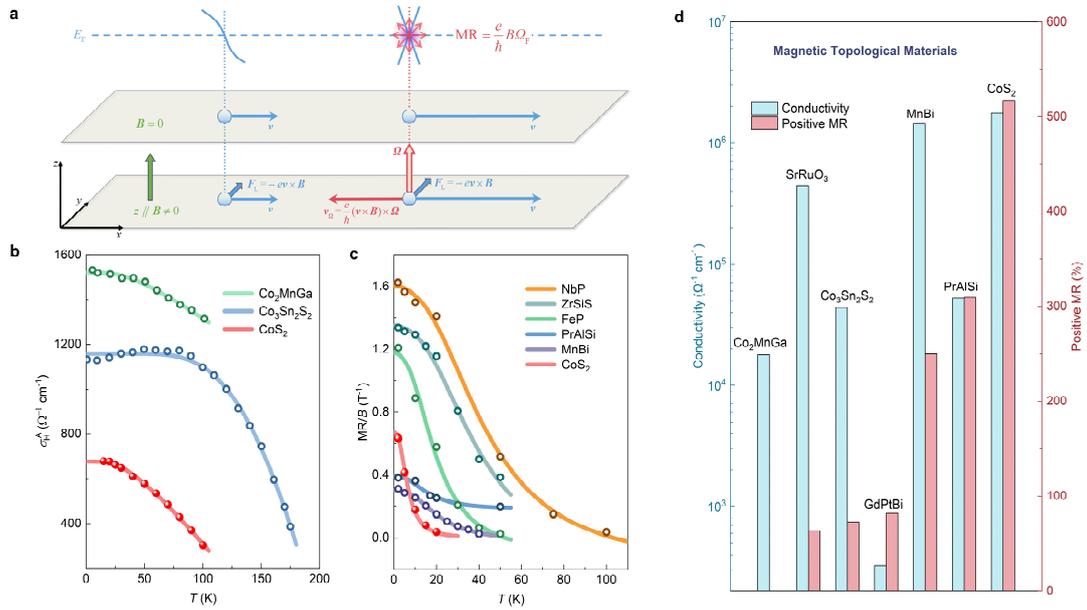

**Fig. 4 | Berry-curvature induced linear positive magnetoresistance (LPMR). a**, Schematic diagram of the intrinsic model. When applying a magnetic field, Weyl fermions gain an additional velocity, which is proportional to the Berry curvature and in the opposite direction to the initial velocity. The slowing Weyl fermions result in a large LPMR phenomenon. **b**, Temperature-dependent AHC in $Co_2MnGa$, $Co_3Sn_2S_2$ and $CoS_2$, and their fitting curves (Eq.(9)) **c**, Temperature-dependent slopes of LPMR in NbP, ZrSiS, FeP, PrAlSi, MnBi, and $CoS_2$, and their fitting curves based on the intrinsic model (Eq.(10)). The data of NbP and ZrSiS are shrunk to match the scale of the other data. Based on the intrinsic model, the slope of LPMR is determined by the average value of the Berry curvature near the Fermi surface (Eq.(5)). The change in slope of LPMR with temperature is consistent with the temperature-dependent Berry curvature. **d**, Conductivity and positive MR in magnetic topological materials. The conductivity at 2 K and the positive MR at 2 K and 9 T are compared. $CoS_2$ possesses the highest conductivity and the largest positive MR among current magnetic topological materials.